\documentclass{article}

\usepackage{PRIMEarxiv}
\usepackage[utf8]{inputenc}
\usepackage[T1]{fontenc}
\usepackage{hyperref}
\usepackage{url}
\usepackage{booktabs}
\usepackage{amsmath,amssymb,amsfonts}
\usepackage{nicefrac}
\usepackage{microtype}
\usepackage{graphicx}
\usepackage[numbers]{natbib}
\usepackage{xcolor}
\usepackage{tabularx}
\usepackage{array}
\usepackage{placeins}

\newcommand{\shorttitle}{Scaling GNNs for Friend Recommendation}

\hypersetup{
	pdftitle={Scaling Graph Neural Networks for Friend Recommendation: Multi-Hash User Embeddings and Temporal Neighbor Sampling},
	pdfsubject={cs.SI, cs.IR, cs.LG},
	pdfauthor={Maksim Utushkin, Andrei Ovsiannikov, Alexander D'yakonov},
	pdfkeywords={friend recommendation, graph neural networks, industrial recommender systems, large-scale graph learning, social graph mining, multi-hash embeddings, temporal neighbor sampling},
}

\providecommand{\Description}[2][]{}

\newcolumntype{Y}{>{\raggedright\arraybackslash}X}
\newcolumntype{Z}{>{\raggedleft\arraybackslash}X}
\newcommand{\compacttablesetup}{%
	\small
	\setlength{\tabcolsep}{3.2pt}%
	\renewcommand{\arraystretch}{1.08}%
}
\newcommand{\codeurl}{\href{https://github.com/makut/VK-GNN}{https://github.com/makut/VK-GNN}}

\begin{document}
	
	\title{Scaling Graph Neural
		Networks for Friend Recommendation: Multi-Hash User Embeddings and
		Temporal Neighbor Sampling%
		\thanks{Accepted at the 35th ACM International Conference on Information and Knowledge Management (CIKM 2026).}}
	
	\author{
		Maksim Utushkin\\
		AI VK\\
		Moscow, Russia\\
		\texttt{mak.utushkin@gmail.com}\\
		\And
		Andrei Ovsiannikov\\
		AI VK\\
		Moscow, Russia\\
		\texttt{andrey.ovsyannikov@vkteam.ru}\\
		\And
		Alexander D'yakonov\\
		AI VK\\
		Moscow, Russia\\
		\texttt{djakonov@mail.ru}\\
	}

	\maketitle
	
	\begin{abstract}
		Friend recommendation is inherently graph-structured: the relevance of a
		potential connection depends on multi-hop social context rather than user
		attributes alone. However, deploying message-passing GNNs on a
		production-scale social graph with hundreds of millions of users and
		tens of billions of edges requires addressing numerous modeling and
		systems challenges. We present a scalable end-to-end GNN ranking system
		for production social graphs, focusing on two design choices that are
		critical in this setting: multi-hash ID embeddings and temporal neighbor
		sampling. Multi-hash embeddings are common for high-cardinality
		features, but industrial GNN systems typically either ignore trainable
		IDs or accept full embedding tables --- exceeding $200$~GB for our graph.
		We integrate multi-hash as the primary node representation, reducing the
		ID-embedding table size by ${>}98\%$ while preserving ranking quality.
		Temporal neighbor sampling is well understood in principle, but existing
		implementations scan full adjacency lists --- a non-starter for users
		with tens of thousands of friends. We implement timestamp-sorted CSR
		storage with binary search, reducing the per-node temporal sampling
		cost from $\mathcal{O}(\deg(v)+k)$ to $\mathcal{O}(\log \deg(v)+k)$. Beyond these components,
		we show that this combination scales and yields measurable production
		impact. On a graph with 194M users and 28B edges, offline ablations
		isolate each design choice's contribution. In an online A/B test, our
		system increases friend additions from recommendations by 16\% and
		unique friend adders by 11.5\% over a strong production baseline. We
		release our framework for distributed training and inference on large
		temporal graphs.
	\end{abstract}

	
	%% =======================================================================
	\section{Introduction}
	\label{sec:intro}
	
	Friend recommendation, often surfaced to users as ``People You May
	Know'' (PYMK), is a core mechanic of any social network: it drives growth and
	connectivity of the social graph and, indirectly, downstream content
	consumption. In production, the task is typically organized as a
	two-stage funnel of candidate generation followed by ranking, with the
	ranker applied across several recommendation surfaces. In this paper we
	focus on the ranking stage and treat it as a learning-to-rank problem
	over user--candidate pairs.
	
	The signal that matters most for friend ranking lives in the structure
	of the social graph: who is connected to whom, how dense the local
	neighborhood around a user is, and how the neighborhoods of two users
	overlap. Graph neural networks (GNNs) are a natural fit, because
	message passing aggregates exactly this multi-hop information into a
	node representation. There is by now a sizable body of work on GNNs
	for recommendation, and several industrial systems have reported
	online wins from GNN-based
	rankers~\cite{lignn2024,grafrank2021,kung2024improving,pinsage2018}.
	
	In our experience, applying a GNN at this scale is less a question of
	which message-passing layer to use and more about a number of
	design decisions that dominate everything else in practice.
	
	Three constraints shape the system in practice. (i)~\emph{Scale}: 
	our graph has roughly 194M nodes and 28B edges, so storage and
	sampling are engineering problems in their own right.
	(ii)~\emph{Weak node content}: profile attributes carry little
	signal, but a trainable per-user ID table is not affordable --- a
	$200\text{M}\times256$ float32 table is itself ${\approx}205$~GB. (iii)~\emph{Dynamic graph}: training on time-ordered
	events while message-passing over the end-of-dataset graph leaks
	future edges; the obvious fix becomes
	bottleneck for hub users with $\deg \sim 10^4$.
	
	\paragraph{Our approach.}
	We tackle the three constraints with two design choices we found
	load-bearing in practice. User IDs enter the GNN through a multi-hash
	layer that maps every user into a shared table of size $B \ll |V|$,
	replacing the infeasible $|V|\times d$ table at the cost of bounded
	hash collisions. Temporal neighbor sampling is implemented over
	timestamp-sorted CSR with a binary-search cutoff, so each sampling
	step becomes barely slower in practice. The
	encoder itself is a GATv2 with two role-specific heads (recipient and
	candidate), trained as binary classification over the impression log.
	
	\paragraph{Contributions.} We make the following contributions:
	\begin{itemize}
		\item We integrate multi-hash ID embeddings as a first-class input
		layer for an industrial GNN ranker, reducing the ID-embedding
		table from ${>}200$~GB to $2$~GB (${<}1\%$) while matching the
		quality of a full $|V|\times d$ table.
		\item We describe a binary-search-based temporal neighbor sampler
		over timestamp-sorted CSR, with $\mathcal{O}(\log d_u + K)$
		per-call cost against $\mathcal{O}(d_u + K)$ for the naive scan
		used by mainstream GNN libraries, eliminating a ${\approx}2.5\times$
		overhead at training time.
		\item We describe the end-to-end pipeline --- CSR storage,
		decoupled CPU sampling and GPU training, offline embedding
		refresh --- that lets the system run on a single 8-GPU host
		despite a $225$~GB graph and 28B edges.
		\item We report offline ablations isolating the contribution of
		each design choice and online A/B results on a production social
		network: $+16\%$ friend additions and $+11.5\%$ unique adders
		over a strong production baseline.
		\item We release the implementation of the training and inference
		framework described in this paper including the components
		needed to train and refresh GNN embeddings over large timestamped CSR
		graphs. The code is available at \codeurl.
	\end{itemize}
	
	%% =======================================================================
	\section{Problem Formulation}
	\label{sec:problem}
	
	\subsection{Graph and task}
	We consider the friendship graph $G = (V, E)$ of a large social
	network: nodes are active users, and an undirected edge $(u, v) \in E$
	records that $u$ and $v$ are mutual friends. The graph is dynamic ---
	each edge carries a timestamp $t_{uv}$ that records when the friendship was formed. Users are reindexed into a contiguous $[0, |V|-1]$
	range during preprocessing, so the same integer ID is used both as a
	CSR index and as a hash input.
	
	Friend recommendations appear on several product surfaces (sidebar,
	dedicated PYMK feed, onboarding), all driven by the same two-stage
	funnel: candidate generation --- product-specific behavioral counters,
	Adamic--Adar~\cite{adamic2003}, and learned retrieval~\cite{walkgnn2024}
	--- followed by ranking. We focus on the ranking stage: given a user
	$u$ and an upstream candidate set $\mathcal{C}_u \subset V$, the model
	scores each pair $(u, v)$, $v \in \mathcal{C}_u$. The system described
	in this paper ranks friend candidates only; it is not part of the
	content-feed ranking stack. Feed-related metrics appear in
	Section~\ref{sec:abtest} solely as downstream indicators of improved
	graph connectivity.
	
	\subsection{Training objective}
	The training signal comes from an interaction log $D$ of recommendation
	impressions, where each record is a tuple $(u_i, v_i, y_i, \tau_i)$. Here, $\tau_i$ is the impression timestamp and $y_i$ is positive
	if the impression resulted in a friend addition, negative on
	no-action impressions and on explicit \texttt{hide} events. The GNN
	score is consumed as a feature by a downstream gradient-boosted
	ranker, which is itself trained on impression-conditioned labels of
	the same form; we adopt this labelling for the GNN as a given, so that
	training and serving operate on the same distribution of pairs.
	The model is trained as binary classification with the standard
	cross-entropy loss
	\begin{equation}
		\mathcal{L} = -\tfrac{1}{|D|}\!\!\sum_{(u,v,y,\tau)\in D}\!\!
		\bigl[\, y \log f(u,v;\tau) + (1-y) \log\bigl(1 - f(u,v;\tau)\bigr) \bigr],
	\end{equation}
	where $f(u,v;\tau)$ uses only information available up to time $\tau$.
	
	%% =======================================================================
	\section{Related Work}
	\label{sec:related}
	
	\noindent\emph{Link prediction in social networks.}
	The academic counterpart is the link prediction problem (LPP),
	introduced for social networks by Liben-Nowell and
	Kleinberg~\cite{liben2003link}: predict which edges will appear in a
	future snapshot. Our setting differs in three ways: positives are
	conditioned on the impression (we predict acceptance, not edge
	formation in the wild), candidates come from an upstream retrieval
	stage rather than all $|V|$ nodes, and the system is
	evaluated on online product metrics rather than a held-out future-edge
	set.
	
	\noindent\emph{Industrial graph-based friend recommendation.}
	Friend recommendation has historically combined structural heuristics
	--- common neighbors, Adamic--Adar~\cite{adamic2003} --- with learned
	embedding-based rankers, and several large social platforms have
	described GNN-style approaches to it. GraFRank~\cite{grafrank2021}
	applies graph attention over a multi-modal friendship graph at
	Snapchat. Another work~\cite{kung2024improving} from the same group
	casts friend recommendation as embedding-based retrieval and focuses on
	the serving stack. SSNet~\cite{ssnet2022} introduces a degree-aware
	re-scaling module for GNN-style friend ranking on Xbox.
	LiGNN~\cite{lignn2024} is an industrial GNN platform at LinkedIn behind
	a number of recommendation surfaces, including friending. Our system
	sits in the same family but focuses on two specific aspects that
	receive less attention in those papers: how to feed hundreds of
	millions of user IDs into the GNN without a giant embedding table, and
	how to keep temporal correctness without paying $\mathcal{O}(\deg(v))$ at every
	sampling step.
	
	\noindent\emph{Graph neural networks and message passing.}
	The mainstream GNN architectures we build on
	--- GCN~\cite{kipf2017}, GraphSAGE~\cite{hamilton2017},
	GAT~\cite{velickovic2018,brody2022} --- share a common message-passing
	template. LightGCN~\cite{he2020lightgcn} simplifies the propagation
	step in collaborative-filtering settings; PinSage~\cite{pinsage2018}
	scales GraphSAGE to web-scale recommendation by sampling neighborhoods
	via random walks. These models are typically described under the
	assumption that nodes either carry rich content features or have a
	trainable per-node embedding. For our task neither assumption is
	attractive: user content features are weak, and per-user embeddings require significant engineering solutions.
	
	\noindent\emph{Compact identifier embeddings.}
	Reducing the size of high-cardinality embedding tables is a recurring
	problem in recommender systems. Feature hashing and hash
	embeddings~\cite{weinberger2009,svenstrup2017} provide a multi-hash
	mechanism that maps a large identifier space into a much smaller shared
	table, trading exact identifiability for memory. Later work generalizes
	this idea, including DHE~\cite{kang2021dhe}, compositional or
	quotient--remainder embeddings~\cite{shi2020compositional}, and
	related compression techniques. In the graph context, position-based
	hash embeddings for GNNs~\cite{kalantzi2021hash} explore similar ideas
	on academic-scale graphs. We treat multi-hash as a first-class input
	layer for an industrial GNN and report its behavior at the scale of
	hundreds of millions of users.
	
	\noindent\emph{Temporal graph learning.}
	Temporal graph models such as TGAT~\cite{xu2020tgat} and
	TGN~\cite{rossi2020tgn} compute node representations as functions of
	time and avoid using future interactions when making predictions. From
	an engineering standpoint, an efficient temporal neighborhood sampler
	is itself a nontrivial component; this is taken up by
	TGL~\cite{zhou2022tgl}, which proposes a temporal-CSR layout and
	parallel sampling routines for dynamic graphs. Our contribution on the
	temporal side is along the same lines: we describe a simple CSR-based
	sampler that uses timestamp sorting plus binary search to bound the
	sampling cost, and we measure its effect on training-time throughput in
	the context of friend ranking.
	
	\noindent\emph{Industrial GNN systems.}
	Beyond modeling, several recent papers describe end-to-end GNN
	platforms for very large graphs: GiGL~\cite{gigl2025} from Snap,
	GraphStorm~\cite{graphstorm2024} from Amazon, and LiGNN~\cite{lignn2024}
	from LinkedIn. These works emphasize neighbor sampling, distributed
	training, and the cost of refreshing representations, all of which are
	also central to our system. Our paper is complementary: rather than a
	full platform paper, we report on two specific design decisions and
	their effect at production scale.
	
	%% =======================================================================
	\section{Method}
	\label{sec:method}
	
	%% (An alternative pipeline overview figure is unused; its TikZ source is
	%%  kept in figures/src/unused/fig_pipeline.tex.)
	
	\subsection{GNN usage}
	We use $L$ stacked GATv2~\cite{brody2022} convolutions as the node encoder. At layer $l$,
	for each neighbor $u$ of vertex $v$, attention scores
	\begin{equation}
		e_{vu}^{(l)} = a^{(l)\top}\operatorname{LeakyReLU}\!\bigl(
		W_s^{(l)} h_v^{(l)} + W_t^{(l)} h_u^{(l)}\bigr)
	\end{equation}
	are mapped to weights with a softmax over set of neighbors of $v$, 
	$\alpha_{vu}^{(l)} = \exp(e_{vu}^{(l)}) / \sum_{k\in\mathcal{N}(v)} \exp(e_{vk}^{(l)})$,
	and the next-layer representation is
	$h_v^{(l+1)} = \sigma\!\bigl(\sum_{u\in\mathcal{N}(v)} \alpha_{vu}^{(l)} W_t^{(l)} h_u^{(l)}\bigr)$.
	
\begin{figure}[t]
		\centering
		\includegraphics[width=0.98\textwidth]{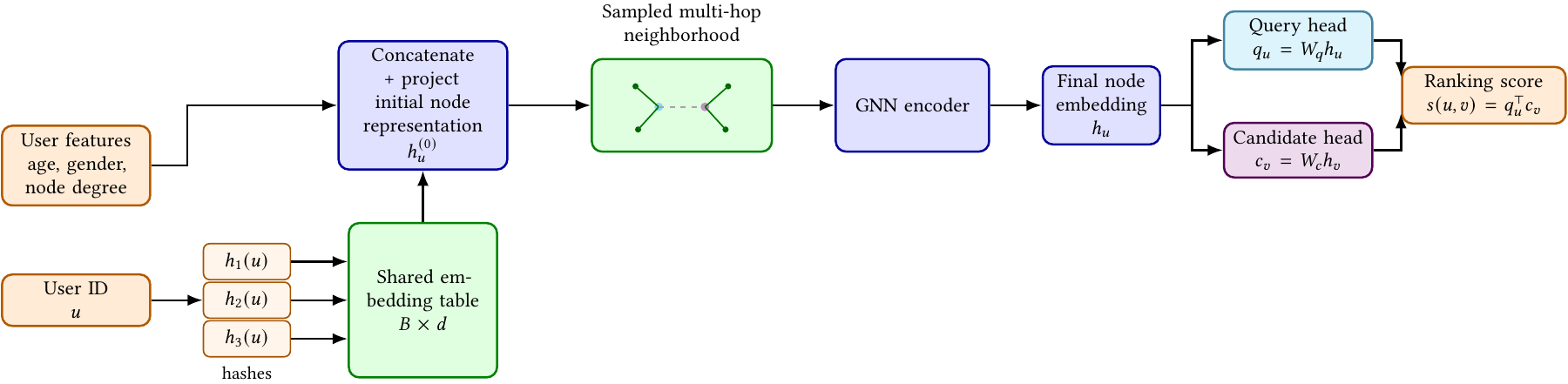}
		\caption{Model overview. User features and multi-hash user-ID
			representations are combined into the initial node embedding, processed
			by a sampled-neighborhood GNN encoder, and projected into separate
			query and candidate representations.}
		\Description[Model pipeline with multi-hash user embeddings and GNN scoring]{
			The figure shows the model pipeline for scoring a user-candidate pair.
			Tabular user features and the user ID are transformed into an initial node
			representation. The user ID is passed through several hash functions, which
			select rows from a shared embedding table. These hash embeddings are
			concatenated, projected, and combined with separately projected tabular
			features. The resulting initial node representation is used as input to a
			GNN encoder over a sampled multi-hop neighborhood. The final node embedding
			is then projected into two role-specific representations: a query
			representation for the user receiving recommendations and a candidate
			representation for the recommended user. The final ranking score is computed
			as the inner product of the query and candidate representations.
		}
		\label{fig:model}
	\end{figure}

	After $L$ layers, the contextual representation $h_v = h_v^{(L)}$
	is projected through two heads,
	$z_u^{\text{q}} = f_{\text{q}}(h_u)$ and
	$z_v^{\text{c}} = f_{\text{c}}(h_v)$, and the pair is
	scored by an inner product
	$s(u, v) = \bigl\langle z_u^{\text{q}},\, z_v^{\text{c}}\bigr\rangle$.
	The two heads let the model treat ``query'' (the user receiving recommendations) and ``candidate''
	roles asymmetrically: the same user appears in both roles in
	training data, and a single representation conflates them.
	Figure~\ref{fig:model} shows the resulting model.
	
	\subsection{Neighbor sampling}
	\label{sec:method-sampling}
	\label{sec:method-temporal}
	The representation of a vertex under a stacked GNN depends on its
	$L$-hop neighborhood. On our graph $L$-hop neighborhoods grow very
	quickly because of high-degree hubs: even $L=2$ can reach tens of
	millions of nodes for a single seed.
	Following standard practice~\cite{hamilton2017,pinsage2018}, we cap the size of the
	computation graph by neighbor sampling: at every hop, we sample a
	bounded number $K$ of neighbors per vertex, so the size of the
	computation graph is bounded by $\mathcal{O}(K^L)$.
	
	Each training example $(u, v, y, \tau)$ carries an event
	timestamp $\tau$, and to avoid information leakage from edges that
	did not yet exist at time $\tau$, we restrict message passing to the
	\emph{temporal} neighborhood
	\begin{equation}
		\mathcal{N}_{<\tau}(u) = \{\, v \in \mathcal{N}(u) : t_{uv} < \tau - \Delta \,\},
	\end{equation}
	where $\Delta \geq 0$ is a safety offset: we exclude edges formed in
	the window $[\tau - \Delta, \tau]$ to be robust to delayed edge
	updates in the production pipeline.
	
	The same cutoff $\tau$ is propagated to every of $L$ hops: when we expand a sampled neighbor $w$ of seed
	$u$, we again restrict to $\mathcal{N}_{<\tau}(w)$, not to a cutoff
	shifted by the edge timestamp $t_{uw}$. The training example, not the
	intermediate edge, fixes the visible history.
	
	The naive way to draw $K$ neighbors from $\mathcal{N}_{<\tau}(u)$ is
	to scan the adjacency list, filter by timestamp, and sample. This is
	what mainstream open-source GNN libraries (DGL~\cite{wang2019dgl}, PyTorch Geometric~\cite{fey2019fast})
	currently do when given a timestamp-aware sampler. For users with
	$\deg(u)$ in the tens of thousands it is too slow.
	
\begin{figure}[tbp]
		\centering
		\includegraphics[width=0.95\textwidth]{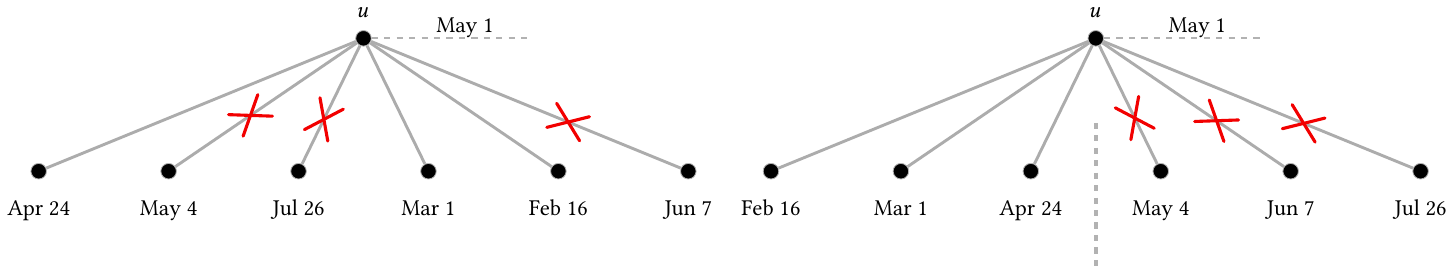}
		\caption{Temporal neighbor sampling. Left: arbitrary-order neighbors require scanning. Right: timestamp-sorted neighbors allow a binary-search split into a valid past prefix and a future suffix.}
		\Description[Comparison of temporal neighbor sampling layouts]{
			A comparison of two temporal neighbor sampling layouts for a user node
			with cutoff time May 1. In the arbitrary-order layout, the user's neighbors
			have unsorted edge timestamps: Apr 24, May 4, Jul 26, Mar 1, Feb 16, and
			Jun 7. Edges after the cutoff time, such as May 4, Jun 7, and Jul 26, are
			invalid and must be found by scanning the adjacency list. In the
			timestamp-sorted layout, neighbors are ordered by edge timestamp: Feb 16,
			Mar 1, Apr 24, May 4, Jun 7, and Jul 26. A binary search finds the cutoff
			between Apr 24 and May 4, splitting the adjacency list into a valid past
			prefix and an invalid future suffix.
		}
		\label{fig:temporal}
	\end{figure}

	We instead store, for each vertex $u$, two coordinated arrays
	$\mathrm{nbrs}_u = (v_1,\dots,v_{d_u})$ and
	$\mathrm{ts}_u = (t_{uv_1},\dots,t_{uv_{d_u}})$, sorted in ascending order of $t_{uv_i}$.
	To draw a temporal sample we binary-search the prefix boundary
	$p = \operatorname{lower\_bound}(\mathrm{ts}_u, \tau - \Delta)$,
	so that $t_{uv_i} < \tau - \Delta$ for all $i < p$, and then sample
	$K$ neighbors uniformly from the valid prefix
	$\bigl(v_1,\dots,v_{p-1}\bigr)$ (Figure~\ref{fig:temporal}).
	In implementation, the naive sampler materializes the
	filtered candidate set by scanning the whole adjacency list, whereas
	the optimized sampler obtains the same candidate prefix by a single
	\texttt{lower\_bound} call and then samples from that prefix.
	
	Let $d_u = |\mathcal{N}(u)|$ and $K$ the requested fanout. The
	binary-search sampler costs $\mathcal{O}(\log d_u + K)$ per call,
	against $\mathcal{O}(d_u + K)$ for the naive scan; the non-temporal
	lower bound is $\mathcal{O}(K)$. The per-call bound relies on a
	one-time preprocessing step: when the CSR is built, the adjacency
	list of every vertex is sorted by edge timestamp using a standard
	comparison sort, at a total cost of
	$\mathcal{O}\bigl(\sum_{u} d_u \log d_u\bigr) =
	\mathcal{O}(|E| \log d_{\max})$ over the whole graph.
	The sorted order is a static property of the snapshot: it is reused
	unchanged by every training epoch and every embedding refresh, and
	sampling itself never re-sorts. For a large graph the per-call gap
	matters in practice: a small number of very high-degree vertices are
	visited often as neighbors, and the $d_u$ versus $\log d_u$ gap
	directly controls sampling throughput. We measure this in
	Section~\ref{sec:exp}.
	
	\subsection{Node representation}
	\label{sec:method-noderepr}
	What to feed into the bottom layer of the GNN is itself a design
	choice. Two approaches dominate: classical papers
	(NGCF~\cite{wang2019ngcf}, LightGCN~\cite{he2020lightgcn}) use a
	learnable embedding table indexed by node ID, while industrial work
	often relies on pre-computed embeddings or content
	features (PinSage~\cite{pinsage2018}, GraFRank~\cite{grafrank2021});
	recent results~\cite{lignn2024} suggest that learnable ID embeddings
	can be highly valuable on industrial graphs. We use both sources: the input
	representation of a user $u$ combines tabular user features and a
	trainable ID-based embedding.

	\paragraph{Tabular features.}
	We use a small set of $m$ tabular user features ($m = 3$ in our
	system): gender, age, and graph degree, where the degree is computed
	in the friendship graph. Gender is treated as a categorical feature,
	while age and graph degree are first quantile-bucketed and then
	treated as categorical features, so that each feature
	$j \in \{1, \dots, m\}$ takes one of $C_j$ discrete values. Each
	feature is represented as a one-hot vector $x_{u,j} \in \{0,1\}^{C_j}$
	and projected through a feature-specific linear layer
	($W_j \in \mathbb{R}^{H \times C_j}$, $b_j \in \mathbb{R}^{H}$) with
	LayerNorm into the hidden dimension $H$ of the encoder ($H = 512$ in
	the base configuration, Table~\ref{tab:training-setup}); the
	per-feature vectors are then summed,
	\begin{equation}
		z_{u,j}^{\text{feat}} = \operatorname{LayerNorm}\!\bigl(W_j x_{u,j} + b_j\bigr)
		\in \mathbb{R}^{H}, \qquad z_{u}^{\text{feat}} = \sum_{j=1}^{m} z_{u,j}^{\text{feat}}
	\end{equation}
	
	\paragraph{Multi-hash ID embedding.}
	The structural signal in our task is commonly stronger than the feature
	signal, so we want a trainable per-user embedding
	on top of features. As discussed in Section~\ref{sec:intro}, a naive
	$|V|\times d$ table usage is limited at our scale, and a
	embedding table would push the serving stack into more complex
	territory.
	
	%% (An alternative multi-hash figure is unused; its TikZ source is kept
	%%  in figures/src/unused/fig_multihash.tex.)
	
	We avoid this with a multi-hash scheme, in the style of hash
	embeddings~\cite{svenstrup2017} and feature
	hashing~\cite{weinberger2009}. A small shared table
	$T \in \mathbb{R}^{B\times d}$ with $B \ll |V|$ is indexed by $k$
	independent hash functions $r_i(u) = h_i(u) \bmod B, \enspace i = 1, \dots, k$.
	The rows $T_{r_1(u)}, \dots, T_{r_k(u)}$ are concatenated and
	projected:
	\begin{align}
		e_u^{\text{hash}} &= T_{r_1(u)} \,\|\, T_{r_2(u)} \,\|\, \cdots \,\|\, T_{r_k(u)} \;\in\; \mathbb{R}^{kd}, \\
		z_u^{\text{emb}}  &= \operatorname{LayerNorm}\!\bigl(W_h\, e_u^{\text{hash}} + b_h\bigr) \;\in\; \mathbb{R}^H.
	\end{align}
	The bottom-layer node representation is then $h_u^{(0)} = z_u^{\text{feat}} + z_u^{\text{emb}}$.
	
	Under a uniform hash family, the probability that two fixed users
	collide on a single hash is $1/B$, so the probability of a full
	collision is $(1/B)^k$, which drops off quickly as
	$k$ grows. Partial collisions on individual hash slots are common and
	acceptable: each partial collision affects one of the $k$
	contributing rows, and the projection $W_h$ can still separate users
	that share some of their hashes.
	
	%% =======================================================================
	\section{Training and Inference System}
	\label{sec:system}
	
	\subsection{Two-stage pipeline}
\begin{figure}[tbp]
		\centering
		\includegraphics[width=\textwidth]{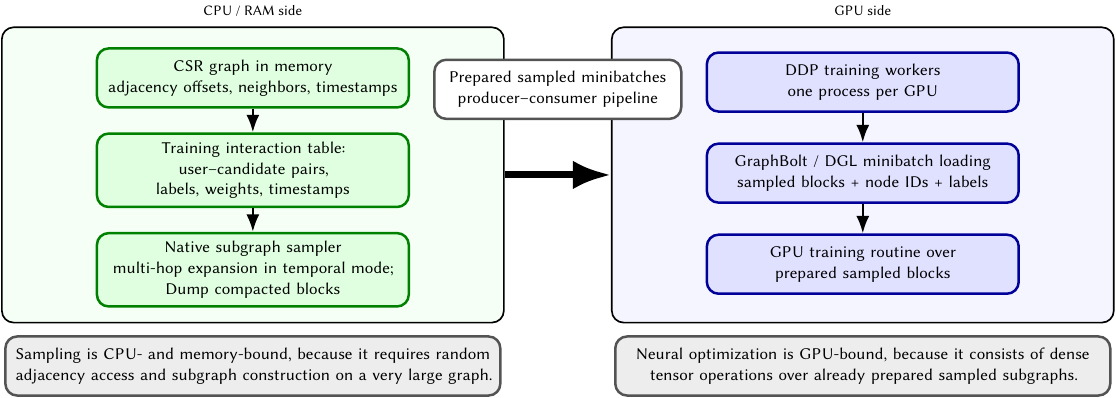}
		\caption{Training system overview. Batch construction is decoupled
			from GPU training: a CPU-side native sampler constructs sampled
			subgraphs, while GPU workers consume prepared minibatches and train the
			GNN model.}
		\Description[Training system split between CPU sampling and GPU training]{
			The figure shows the training pipeline split into a CPU/RAM side and a GPU
			side. On the CPU/RAM side, the system keeps the CSR graph in memory,
			including adjacency offsets, neighbor IDs, and timestamps, together with
			the training interaction table containing user-candidate pairs, labels,
			weights, and timestamps. A native subgraph sampler performs temporal
			multi-hop expansion over the CSR graph and serializes compacted sampled
			blocks into prepared minibatches. These minibatches are passed through a
			producer-consumer pipeline to the GPU side. On the GPU side, one DDP
			training worker runs per GPU, loads the sampled blocks and node IDs through
			GraphBolt or DGL minibatch loading, and performs the GNN training step.
			The diagram emphasizes that subgraph construction is CPU- and
			memory-bound, while neural optimization over prepared sampled blocks is
			GPU-bound.
		}
		\label{fig:trainsys}
	\end{figure}

	Training a GNN has two qualitatively different
	workloads: CPU-bound batch construction (multi-hop neighborhood
	sampling, temporal filtering,
	dominated by random CSR access and memory bandwidth) and GPU-bound
	training. We decouple them. A sampler (Java/C++ with Python
	binding) consumes the CSR graph and batch seeds and
	serializes minibatches into a bounded queue; each contains the sampled
	message-passing blocks (per-layer bipartite subgraphs), compacted node IDs for feature lookup,
	original user IDs for the multi-hash lookup, and per-example labels. A Python trainer on
	PyTorch and DGL/GraphBolt~\cite{wang2019dgl} reads from the queue and runs
	the optimizer. The two stages scale independently --- CPU workers
	absorb sampling backlog, the prefetch buffer absorbs GPU stalls
	(Figure~\ref{fig:trainsys}).
	
	\subsection{Hardware and runtime}
	\label{sec:hardware}
	Both training and inference run on a single host with 8 NVIDIA A100 80~GB GPUs, 512~GB
	of system RAM, and 64 CPU cores. The CSR graph and the tabular feature tables sit
	on the CPU side; the multi-hash table, model parameters, optimizer
	state, and activations sit on the GPU side. Under the base
	configuration (Section~\ref{sec:exp-setup}) the multi-hash table
	is about $2$~GB, and the full GPU-side state fits comfortably into
	one device and is replicated across DDP ranks.
	
	\subsection{Graph storage}
	\label{sec:graph-storage}
	The graph is stored separately from the interaction log.
	After reindexing users to $[0, |V|-1]$, the graph is held in a CSR
	layout:
	\begin{itemize}
		\item \texttt{indptr} --- the per-vertex offsets into the neighbor
		array;
		\item \texttt{indices} --- the concatenated neighbor IDs;
		\item \texttt{timestamps} --- timestamps aligned with
		\texttt{indices}.
	\end{itemize}
	Maximum value in \texttt{indptr} is bounded by $|E|-1$, and in
	\texttt{indices} by $|V|-1$. With $|V|$ of order a few hundred million,
	the neighbor IDs fit in 32-bit integers, whereas
	\texttt{indptr} needs 64-bit because the total number of edges exceeds
	$2^{32}$. Using 32-bit \texttt{indices} cuts the memory footprint of
	the large array in half. \texttt{timestamps} are kept in 32-bit integers as well (Unix
	seconds with offset).
	
	For the production graph, this gives a concrete CSR footprint of
	\begin{itemize}
		\item \texttt{indices}: $28\cdot 10^{9}\cdot 4~\text{B} \approx 112~\text{GB}$;
		\item \texttt{timestamps}: $28\cdot 10^{9}\cdot 4~\text{B} \approx 112~\text{GB}$;
		\item \texttt{indptr}: $(|V|+1)\cdot 8~\text{B} \approx 1.5~\text{GB}$,
	\end{itemize}
	for a total of $\approx 225$~GB. This fits in the RAM of the host (Section~\ref{sec:hardware}), which is the reason we can
	keep the entire graph in process.
	
	For temporal sampling, neighbors inside each vertex's adjacency list
	are sorted in ascending timestamp order during graph construction (see
	Section~\ref{sec:method-temporal}). The same CSR format is reused at
	inference.
	
	\subsection{Inference and embedding refresh}
	At inference time we reuse the same neighborhood expansion. For a
	batch of users, we materialize sampled local subgraphs and run the
	trained encoder, producing user embeddings consumed by the production ranker.
	
\begin{figure}[tbp]
		\centering
		\includegraphics[width=\textwidth]{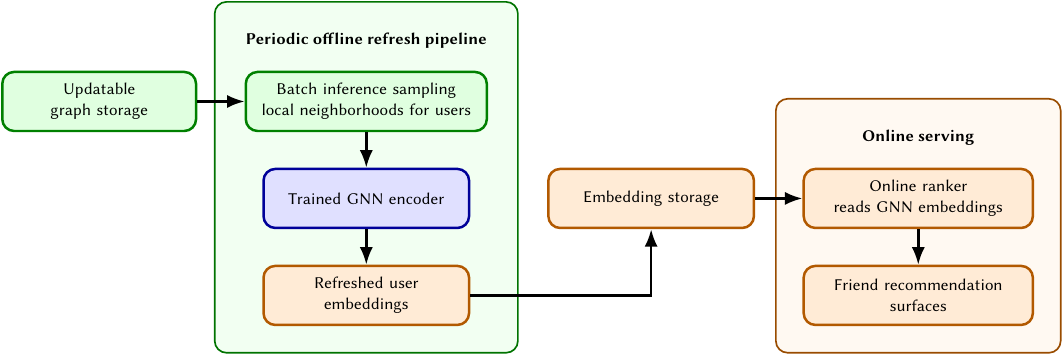}
		\caption{Inference and embedding refresh. The periodic offline refresh
			pipeline samples local neighborhoods from the updatable graph storage,
			runs the trained GNN encoder, and writes refreshed user embeddings to
			the embedding storage. The online ranker reads these embeddings during
			serving.}
		\Description[Inference and embedding refresh pipeline]{
			The figure shows a two-stage pipeline that separates offline GNN embedding
			refresh from online friend-recommendation serving. In the offline refresh
			stage, an updatable graph storage block provides graph data to a batch
			inference sampling block, which samples local neighborhoods for users.
			These sampled neighborhoods are passed to a trained GNN encoder, which
			produces refreshed user embeddings. The embeddings are then written to an
			embedding storage block. In the online serving stage, the online ranker
			reads the precomputed embeddings from embedding storage and uses them to
			serve friend recommendation surfaces. The diagram emphasizes that graph
			sampling and GNN encoding are performed offline, while the online ranker
			only consumes already computed embeddings.
		}
		\label{fig:inference}
	\end{figure}

	The refresh policy is dictated by the dependency structure of a GNN.
	Unlike a two-tower model, where a user embedding depends only on the
	user's own features, a GNN representation depends on a sampled
	neighborhood, so a single new edge can affect many
	embeddings. We do not track such cascades online. Instead, the
	encoder periodically recomputes representations for a large active
	subset. Section~\ref{sec:exp} reports
	latency and refresh-cost numbers. Figure~\ref{fig:inference} shows the
	flow. The model is retrained from scratch in the same periodic manner; we discuss cold-start implications in Section~\ref{sec:cold-start}.
	
	%% =======================================================================
	\section{Experiments}
	\label{sec:exp}
	
	\subsection{Setup}
	\label{sec:exp-setup}
	We use a snapshot of the production friendship graph and the matching
	impression log; Table~\ref{tab:dataset} summarizes the scale.
	Interactions are split temporally on the impression timestamp, with
	the last three years used for training. The label is binary, as
	defined in Section~\ref{sec:problem}.
	
	\begin{table}[t]
		\centering
		\caption{Dataset statistics.}
		\label{tab:dataset}
		\begin{tabular}{lrrrrr}
			\toprule
			Statistic & $|V|$ & $|E|$ & Train interactions & Test interactions \\
			\midrule
			Value     & 194M          & 28B                 & 1.4B               & 25M \\
			\bottomrule
		\end{tabular}
	\end{table}
	
	Unless noted otherwise, every model is trained with the configuration
	in Table~\ref{tab:training-setup}; ablations change one parameter at a
	time and keep the rest fixed.
	
	\begin{table}[t]
		\centering
		\caption{Base training configuration. All ablations are run with one
			parameter changed at a time; the rest are held fixed at the values
			below.}
		\label{tab:training-setup}
		\compacttablesetup
		\begin{tabularx}{\columnwidth}{@{}p{0.34\columnwidth}Y@{}}
			\toprule
			Component & Setting \\
			\midrule
			GNN encoder            & GATv2, $L = 2$ layers, $8$ attention heads \\
			Hidden dimension       & $H = 512$ \\
			Head output dimension  & $128$ (query, candidate) \\
			Multi-hash slots       & $k = 3$ \\
			Shared hash table      & $B = 2^{21}$ rows, dim $d = 256$ \\
			Hash function          & 64-bit integer multiplicative hashing \\
			Per-hop fanout         & $K = 30$ \\
			Temporal sampling      & binary-search cutoff with shift $\Delta = 30\,\mathrm{min}$ \\
			\midrule
			Optimizer              & Adam, learning rate $1.5{\cdot}10^{-3}$ \\
			Batch size             & $1024$ user--candidate pairs per GPU \\
			Parallelism            & DDP on 8 A100 GPUs \\
			Stopping rule          & early stopping on validation ROC-AUC\\
			\bottomrule
		\end{tabularx}
	\end{table}
	
	\subsection{Baselines}
	\label{sec:exp-baselines}
	We compare against three baselines that span the range from a
	non-learned popularity prior to the previous production ranker:
	\begin{itemize}
		\item \textbf{Top-pop.} Candidates are ranked by their in-degree in
		the friendship graph. The model has no learnable parameters and
		serves as a sanity-check lower bound.
		\item \textbf{MF.} A matrix factorization model trained on the same
		interaction log as the GNN.
		\item \textbf{WalkGNN.} The previous production solution in our company,
		described in~\cite{walkgnn2024}. It scores user--candidate pairs by
		aggregating GNN-based relevance estimates over local ego-net contexts.
		We use it as the strongest production baseline. In contrast, our model precomputes user embeddings offline and reduces online pair scoring
		to an inner product, making the serving path simpler and cheaper
		than constructing ego-subgraphs per scored pair.
	\end{itemize}
	For the proposed model we report the best configuration from the
	ablations below.
	
	\subsection{Offline evaluation}
	\label{sec:offline}
	Table~\ref{tab:main} reports per-user ROC-AUC for all four models.
	Each step adds signal: MF captures collaborative patterns,
	WalkGNN adds local ego-network structure, and our model adds the
	largest increment by extending message passing to the full multi-hop
	neighborhood --- a step enabled by the multi-hash and temporal-sampling
	choices ablated in Section~\ref{sec:ablations}.
	
	\begin{table}[t]
		\centering
		\caption{Offline ranking quality on the held-out test
			interactions, measured as per-user ROC-AUC (higher is better).}
		\label{tab:main}
		\begin{tabular}{lcccc}
			\toprule
			& Top-pop & MF     & WalkGNN & Ours   \\
			\midrule
			ROC-AUC & 0.5050  & 0.5316 & 0.5572 & \textbf{0.6278} \\
			\bottomrule
		\end{tabular}
	\end{table}
	
	\subsection{Ablations}
	\label{sec:ablations}
	We ablate each of the design choices in turn. The backbone is described in Table~\ref{tab:training-setup}; each ablation varies one component at a time.
	
	\subsubsection{Input representations}
	\label{sec:abl-input}
	We isolate the contributions of tabular features and the multi-hash
	ID embedding, and additionally compare against a full $|V|\times d$
	trainable embedding table as a high-capacity reference point. The full table
	does not fit on a single host and is implemented offline via
	TorchRec~\cite{ivchenko2022torchrec} row-sharding across the 8 training
	GPUs; it is not a serving option but quantifies the quality cost of the
	multi-hash compression.
	
	\begin{table}[t]
		\centering
		\caption{Effect of the ID-representation scheme on offline ranking
			quality (per-user ROC-AUC).}
		\label{tab:abl-hash}
		\compacttablesetup
		\begin{tabularx}{\columnwidth}{@{}Y>{\raggedleft\arraybackslash}p{0.21\columnwidth}>{\raggedleft\arraybackslash}p{0.17\columnwidth}@{}}
			\toprule
			ID scheme                                   & Embedding table size                & ROC-AUC          \\
			\midrule
			Features only                               & ---                                 & 0.5244           \\
			Multi-hash IDs only (no features)           & $2$~GB                                & 0.5997           \\
			Features + full table                       & $202.88$~GB & 0.6246           \\
			Features + multi-hash IDs \emph{(ours)}     & $2$~GB                                & \textbf{0.6278}  \\
			\bottomrule
		\end{tabularx}
	\end{table}
	
	Two observations stand out. First, the structural signal alone
	outperforms the tabular signal alone by a wide margin, confirming
	that for friend ranking the graph carries more than the profile.
	Second, multi-hash matches the full $|V|\times d$ table while using
	${<}1\%$ of its memory; we hypothesize that the bounded sharing
	induced by multi-hash acts as an implicit regularizer that prevents
	overfitting on individual user rows.
	
	\subsubsection{Multi-hash table size}
	\label{sec:abl-hashsize}
	Table~\ref{tab:abl-hashsize} sweeps the shared-table size $B$ from
	$2^{17}$ to $2^{22}$ rows with $k=3$ and $d=256$ fixed.
	
	\begin{table}[t]
		\centering
		\caption{Effect of the shared-table size $B$ on offline ranking
			quality (per-user ROC-AUC) and on the memory required.}
		\label{tab:abl-hashsize}
		\compacttablesetup
		\begin{tabular}{@{}l*{6}{r}@{}}
			\toprule
			$B$               & $2^{17}$ & $2^{18}$ & $2^{19}$ & $2^{20}$ & $2^{21}$ & $2^{22}$       \\
			\midrule
			Hash table memory & $128$~MB & $256$~MB & $512$~MB & $1$~GB   & $2$~GB                 & $4$~GB         \\
			ROC-AUC           & $0.5952$ & $0.6044$ & $0.6092$ & $0.6184$ & $0.6278$               & $\mathbf{0.6310}$ \\
			\bottomrule
		\end{tabular}
	\end{table}
	
	Quality grows monotonically with $B$ and has not yet flattened at
	$2^{22}$. We pick $B = 2^{21}$ for the production configuration: at
	$2^{22}$ the table doubles in size while quality changes only
	marginally.
	
	\subsubsection{Temporal sampling}
	\label{sec:abl-temporal}
	Switching off temporal sampling lets the model aggregate over edges
	that did not yet exist at prediction time --- a form of label leakage.
	Table~\ref{tab:abl-temporal} reports both the quality impact and the
	per-batch sampling cost.
	
	\begin{table}[htbp]
		\centering
		\caption{Effect of temporal neighbor sampling on offline quality
			and on per-batch sampling cost. Sampling
			time is wall-clock cost per minibatch under the base
			configuration.}
		\label{tab:abl-temporal}
		\compacttablesetup
		\begin{tabularx}{\columnwidth}{@{}Y>{\raggedleft\arraybackslash}p{0.20\columnwidth}>{\raggedleft\arraybackslash}p{0.25\columnwidth}@{}}
			\toprule
			Sampler                                & ROC-AUC           & Sampling time                       \\
			\midrule
			Non-temporal                           & $0.5907$          & $581$~ms \\
			Temporal, naive scan                   & $0.6278$ & $1473$~ms \\
			Temporal, binary search \emph{(ours)}  & $\mathbf{0.6278}$ & $595$~ms \\
			\bottomrule
		\end{tabularx}
	\end{table}
	
	The non-temporal model trains on a graph that includes
	post-impression edges and is $0.0371$ ROC-AUC below the temporal
	version, quantifying the leakage. Both temporal samplers draw from
	the same $\mathcal{N}_{<\tau}(u)$ and yield identical quality; the
	binary-search sampler reaches it at essentially the same per-batch
	cost as the non-temporal one, while the naive
	scan is ${\approx}2.5\times$ slower.
	
	\subsection{System scalability}
	\label{sec:scalability}
	Table~\ref{tab:scalability} reports the operational numbers that
	matter for an industrial deployment, all measured under the base
	configuration of Table~\ref{tab:training-setup}.
	
	\begin{table}[htbp]
		\centering
		\caption{End-to-end system numbers.}
		\label{tab:scalability}
		\footnotesize
		\setlength{\tabcolsep}{3.0pt}
		\renewcommand{\arraystretch}{1.04}
		\begin{tabularx}{\columnwidth}{@{}Y>{\raggedleft\arraybackslash}p{0.31\columnwidth}@{}}
			\toprule
			\multicolumn{2}{l}{\emph{Training time}}                                                  \\
			\midrule
			GPU step (forward + backward per batch)  & $927$~ms           \\
			Wall-clock time per epoch                & $43.4$~h            \\
			Iterations to early stopping                 & $252$~k              \\
			Total training time to convergence       & $63$~h            \\
			\midrule
			\multicolumn{2}{l}{\emph{Inference and memory}}                                            \\
			\midrule
			Embedding computation for all 194M users & $6.57$~h            \\
			CSR graph, CPU side                      & $225$~GB                                      \\
			Tabular feature tables, CPU side         & $2.18$~GB           \\
			Multi-hash table, GPU side               & $2$~GB                                        \\
			Model parameters + Adam state, GPU side  & $6.03$~GB           \\
			\bottomrule
		\end{tabularx}
	\end{table}
	
	The graph occupies $225$~GB but the trainable parameter
	set fits on a single GPU --- a direct
	consequence of the multi-hash design.
	
	\subsection{Online A/B test}
	\label{sec:abtest}
	The model trained with the configuration in
	Table~\ref{tab:training-setup} was deployed as a ranking feature in
	the production friend recommendation pipeline. The treatment group
	used a ranker enriched with the GNN score; the control group used the
	previous production ranker. Table~\ref{tab:abtest} summarizes the online effects.
	
	\begin{table}[htbp]
		\centering
		\caption{Online A/B test results against the previous production
			ranker.}
		\label{tab:abtest}
		\compacttablesetup
		\begin{tabularx}{\columnwidth}{@{}Y>{\raggedleft\arraybackslash}p{0.25\columnwidth}@{}}
			\toprule
			Metric & Relative change \\
			\midrule
			Friend additions from recommendations & $+16.0\%$ \\
			Unique users adding a recommended friend & $+11.5\%$ \\
			Total time spent in content feed & $+0.28\%$ \\
			Ranker p50/p90/p99 latency & no regression \\
			\bottomrule
		\end{tabularx}
	\end{table}
	
	The experiment was run on production traffic for two weeks. The two
	direct friending metrics improved by $+16.0\%$ and $+11.5\%$,
	respectively, both statistically significant at $p < 0.01$. We also
	observed a statistically significant $+0.28\%$ lift in total time
	spent in the content feed, which we consider as a purely downstream effect: users form more connections and therefore get more content.
	
	Because the GNN signal is delivered as precomputed embeddings rather
	than computed at request time, ranker latency is within noise compared to the control group.
	
	\FloatBarrier
	
	%% =======================================================================
	\section{Discussion and Limitations}
	\label{sec:discussion}
	
	\subsection{Cold start and new users}
	\label{sec:cold-start}
	The multi-hash scheme is tied to the user cohort seen during
	training: a user who joined after the snapshot, or who was inactive at
	training time, has no incident edges in the
	graph and gets no learning signal. We
	handle this by retraining periodically on a fresh graph; between runs,
	such users are served by upstream candidate generation and the GNN
	re-ranker falls back to a representation whose only ID-side content
	comes from incidental hash collisions (the features-only ablation in
	Section~\ref{sec:abl-input} bounds that fallback). The retraining
	cadence sets the size of the cold window. Several directions
	are compatible with our system: inductive aggregation over the partial
	neighborhood (GraFRank~\cite{grafrank2021}), continuous-memory updates
	between runs (TGN~\cite{rossi2020tgn}), graph densification with
	synthetic edges (LiGNN~\cite{lignn2024}), and low-rank warm-start
	folding-in~\cite{yusupov2025warmstart}.
	
	\subsection{Non-real-time serving}
	\label{sec:non-realtime}
	A GNN embedding depends on an $L$-hop sampled neighborhood,
	so producing one online requires the same multi-hop expansion the
	trainer runs on the CPU --- which does not fit the latency budget of
	a ranker call (two-tower models avoid this, since each side's
	representation depends only on its own features). We therefore keep
	the GNN offline: embeddings are refreshed on a fixed schedule
	and exposed to the online ranker as features without
	raising ranker p99 latency. The price is staleness between refreshes. For
	friend recommendation this is favourable, since friendship is a
	slowly-changing signal compared to feed-style engagement; on surfaces with within-session interest
	shift, a continuously-updated memory module~\cite{rossi2020tgn} at
	higher per-request cost would be a better fit.
	
	\subsection{Beyond the friendship graph}
	\label{sec:hetero}
	Both design choices transfer to heterogeneous graphs, where many recommender deployments live. The temporal sampler is
	unchanged: each edge type carries its own CSR sorted by timestamp,
	and a cutoff query is a binary search per edge type with the same $\mathcal{O}(\log d_u + K)$ cost. The multi-hash representation
	transfers more selectively --- it need not be applied to every node
	type. In a user--item bipartite graph, for instance, items
	typically have strong content features while
	users have weak profile attributes, so it is natural to feed content
	features to items and multi-hash to users.
	
	%% =======================================================================
	\section{Conclusion}
	\label{sec:conclusion}
	
	We described a production GNN ranking system for friend recommendation
	on a huge graph with 194M users and 28B edges. The two design choices we focused on --- a multi-hash
	representation of user IDs as the primary input layer of the GNN, and a
	binary-search-based temporal neighbor sampler --- are relatively simple,
	but together they cover the two parts of the problem that we found
	hardest to get right in practice. In an online A/B test, the system
	improved friend additions from recommendations by $+16\%$ and also
	increased total time spent in the content feed by $+0.28\%$. By releasing
	the framework --- including the native temporal neighbor sampler, the multi-hash embedding layer, and the training and inference pipeline --- we hope to make these components
	useful to other teams building GNN rankers at industrial scale.
	
	%% =======================================================================
	
	\section*{GenAI Usage Disclosure}
	The authors used generative AI tools, including ChatGPT, Claude, and
	DeepSeek, during manuscript preparation for language editing,
	alternative phrasings, structural feedback, and assistance with LaTeX
	snippets. These tools were also used to assist with refactoring and
	adapting selected runtime components for open-source release, and with
	writing parts of the training-launch scripts. The core framework
	implementation was written manually by the authors. Generative AI tools
	were not used to generate experimental results, labels, production
	metrics, or the core technical claims of the paper. All AI-assisted text,
	code suggestions, and refactoring suggestions were reviewed, edited,
	tested where applicable, and verified by the authors, who take full
	responsibility for the final content.
	
	\bibliographystyle{IEEEtran}
	\bibliography{refs}
	
\end{document}